
%
%
%
%
%
%

\long\def\UN#1{$\underline{{\vphantom{\hbox{#1}}}\smash{\hbox{ #1}}}$}
\def\NL{\hfill\break}
\def\NI{\noindent}
\magnification=\magstep 1
\overfullrule=0pt
\hfuzz=16pt
\voffset=0.0 true in
\vsize=8.8 true in
   \def\NP{\vfil\eject}
   \baselineskip 20pt
   \parskip 6pt
   \hoffset=0.1 true in
   \hsize=6.3 true in
\nopagenumbers
\pageno=1
\footline={\hfil -- {\folio} -- \hfil}
\headline={\ifnum\pageno=1 \hfill September 1992 \fi}

\hphantom{AA}

\hphantom{AA}

\centerline{\bf DISCRETE TO CONTINUOUS-TIME CROSSOVER DUE TO
ANISOTROPY}
\centerline{\bf IN DIFFUSION-LIMITED TWO-PARTICLE
ANNIHILATION REACTIONS}

\vskip 0.4in

\centerline{\bf Vladimir~Privman}

\vskip 0.2in

{\NI}{\sl{Department of Physics, Clarkson University, Potsdam, New
York \ 13699--5820, USA}}

\vskip 0.4in

\NI {\bf PACS:}$\;$ 05.70.Ln, 68.10.Jy, 82.20.Mj

\NI {\bf Key Words:}$\;$ Diffusion-Limited Reactions, Anisotropic
Hopping, \NL \hphantom{\bf Key Words:}$\;$ Exact Solutions, Continuum Limit

\NI {\bf Running Title:}$\;$ Anisotropy in Diffusion-Limited
Reactions

\vskip 0.4in

\centerline{\bf ABSTRACT}

Diffusion-limited reaction A$+$A$\to$inert with anisotropic hopping
on the $d=1$
lattice, is solved exactly for a simultaneous updating, discrete time-step
dynamics.
Diffusion-dominated processes slow down as the anisotropy
increases. For large times
or large anisotropy, one can invoke the appropriate continuum limits.
In these limits
the effects of the anisotropy on variation of particle density can be
absorbed in
time rescaling. However, in other regimes, when the discreteness of
the time steps is
nonnegligible, the anisotropy effects are nontrivial, although they are
always quite
small numerically.

\NP

\NI \UN{\bf 1. \ Introduction}

Recent studies of fluctuation effects in reaction-diffusion systems
have
been focused largely on exact or asymptotic continuum-limit
analyses of
low-dimensional, $d=1$ and $d=2$ models of the simplest reactions
such as A$+$A$\to$inert,
A$+$A$\to$A, etc.~[1-12]. Particle input, fragmentation, and
multiparticle reactions have been
considered as well [4,8,12-13]. Fluctuation effects are significant in
low dimensions;
diffusion-limited reactions then show variation of the particle
density at large times which is
different from the predictions of the chemical rate-equation
approach (mean-field theory). In
this work we initiate studies of the effects of anisotropy in the
hopping rates on lattice
reaction-diffusion models. Specifically, we derive exact results for
the reaction
A$+$A$\to$inert, in $d=1$.

Systems of hard-core particles show interesting novel density
fluctuations associated with
hopping rate anisotropy, such as shock-waves, etc. Recent studies
[14-16] have yielded a variety
of information pertinent to these phenomena, with most explicit
results limited to $d=1$ models.
We note however that the few simplest, \UN{\sl solvable\/}  reaction
schemes in $d=1$ involve
instantaneous reactions such as two-particle annihilation. These
reactions are termed
diffusion-limited. The hard core property cannot be combined with
annihilation on encounter
unless the latter occurs with probability $<1$. The partial-reaction
probability models can only
be studied numerically [10,13]. Added hopping anisotropy may also
have a profound effect on
reactions in nonuniform systems with, e.g., point-like sources of
particles, sinks, walls, etc.
Investigation of these effects is, however, outside the scope of the
present study.

Here we focus on the third interesting effect of the hopping rate
anisotropy: slowing down of
the diffusion. Indeed, let us consider a single particle, hopping on a
line with the rate $H$ per
unit time, and with the fixed step length $\ell$ to the right and to the
left. Let the hopping rate
to the right be $(1+a)H/2$, while that to the left be $(1-a)H/2$. Here
we assume
$-1 \leq a \leq 1$. In the long-time regime when the continuum
description can be used, the
particle diffusion constant $D(a)$ will be

$$ D (a) = (1-a^2) \ell^2 H /4 = (1-a^2) D(0)   \; .  \eqno(1.1) $$

\NI Of course the diffusion is superimposed on the drift with velocity
$a\ell H$.
Note however that while the time scale of the drift is set by the
original time variable $t$, the
time scale of the diffusional spread is set by the dimensionless
variable

$$ \tau (a) = 4 D(a) t / \ell^2 = (1-a^2) t   H  \; ,  \eqno(1.2) $$

\NI where the coefficient 4 was introduced for convenience. Thus,
diffusion-dominated
processes get slowed down as the anisotropy increases.

Our reaction-diffusion model for the $d=1$ reaction A$+$A$\to$inert,
will be defined in
Section 2, within the cellular-automaton, simultaneous updating
formulation [11], with discrete
time steps $\Delta t =1$. However, we will demonstrate (Section 4)
that the model
approaches the continuum-time reaction-diffusion kinetics when the
``natural'' time steps,
$\Delta \tau = (1-a^2) H \Delta t $, vanish in the extreme anisotropy
limit $|a| \to 1$. Exact solution
will be obtained for all $a$ values, in Section 3. The $a=0$ case
corresponds to the isotropic
reaction solved in the cellular-automaton variant in [11]. The $|a|=1$
limiting solution turns out
(Section 4) to be identical with the result obtained in the isotropic
case for the continuum-time
Glauber-dynamics variant of the reaction A$+$A$\to$inert, solved in
[5,9]. Section 4
also contains a summarizing discussion.

\

\NI \UN{\bf 2. \ Definition of the Model}

We consider particles hopping on a linear $x$-lattice of spacing
$\ell$. The dynamics is defined
to have the evolution in time steps $\Delta t = 1$. At a given time $t
= 0,1,2, \ldots$, we
assume that the reacting particles occupy, with certain average
density $\rho (t)$, odd lattice
sites, $x/ \ell = \pm 1, \pm 3, \ldots$, at even times ($t
=0,2,4,\ldots$), and even lattice sites
($x/ \ell = 0, \pm 2, \ldots$) at odd times $t =1,3,\ldots\;$. There can
be 0 or 1 particles at
each site (of the relevant sublattice). At each time step $t \to t+1$,
each of the particles
randomly hops one lattice spacing to the right  $(+\ell \hat x )$ or to
the left $(-\ell \hat x )$,
with respective probabilities $(1+a)/2$ and $(1-a)/2$. The hopping
events are independent.
However, at those sites which at time $t+1$ received 2 particles, the
pairs ``annihilate,'' i.e.~the
occupation number is immediately reduced to 0. It is obvious that
this rule decouples the even and
odd sublattices at alternative time steps, --- the reason for our
restriction to the even/odd
sublattices as described earlier.

Quantities such as the particle density per site $\rho (t)$ are
calculated by
averaging not only over the hopping direction choices but also over
the initial particle
distribution at time $t=0$. In order to have an exactly solvable
model, here as for the isotropic
case $a=0$ [9,11], one has to consider a particular choice of the initial
particle distribution:
randomly occupied (odd) sites with the average initial density $\rho
(0) = 1/2$
(per odd-sublattice site).

The root to exact solutions of annihilation models A$+$A$\to$inert,
involves the
transformation [5,9,11] from the hopping particle dynamics to the
dual-lattice stochastic spin
dynamics. In order to introduce the dual lattice systematically, let us
describe the original
particle system in the following somewhat unusual way: the
occupancy variables $m_j (t)$
will take values $-1$ or 1, for occupied sites $x=j\ell $ or empty sites,
respectively. Thus
the particle density is

$$ \rho (t) = \langle [1-m_j(t) ] /2 \rangle \; . \eqno(2.1) $$

\NI Note that the density does not depend on $x = j \ell $ because
the initial
conditions and stochastic hopping rules, --- average over both being
denoted by
$\langle \ldots \rangle$, --- are translationally invariant.

We now introduce the dual-lattice spins $\sigma_j (t) = \pm 1$, on
the even sublattice
for even times, and on the odd sublattice for odd times. Thus, these
spins are located
in the interstices $x=j\ell$ between the allowed particle-location sites
at $x=(j\pm
1)\ell $. Formally, the spin values are defined by the infinite
products

$$ \sigma_j (t) = \prod_{i=0}^\infty m_{j+2i+1} (t) \; , \eqno(2.2) $$

\NI although with some care in definitions, products up to some fixed
$ m_J (t) $ could be
used as well. This definition implies

$$m_j (t) = \sigma_{j-1}(t) \sigma_{j+1}(t) \; , \eqno(2.3) $$

\NI so that the density is given by

$$ 2 \rho (t) = 1-\langle \sigma_{j-1} (t) \sigma_{j+1} (t)  \rangle = 1
- G_2 (t) = G_0 (t) -
G_2 (t) \; , \eqno(2.4) $$

\NI where we define the two-point spin-spin correlation function

$$ G_n (t) = \langle \sigma_i (t) \sigma_{i+n} (t)  \rangle  \; ,
\eqno(2.5) $$

\NI which only depends on the separation $|n|$ due to translational
invariance. We will consider
this function for $n=0,2,4,\ldots\;$.

While the stochastic dynamics of the original occupancy variables
$m_i (t)$ is nonlinear, it
reduces to a linear dynamical rule for the dual spins. This can be
shown directly by first
introducing the particle dynamics. However, the steps involved are
cumbersome. A simpler
approach is to define the linear spin dynamics first and check that
the particle hopping and
annihilation are correctly represented by it. Thus, let us introduce
random variables $\xi_i (t)$
taking on values 1 or 0, with respective probabilities $(1-a)/2$ and
$(1+a)/2$, where $a$ is the
hopping bias. Then the dynamical rule is

$$ \sigma_i (t+1) = \left[ 1 - \xi_i (t) \right]  \sigma_{i-1} (t)
 + \xi_i (t) \sigma_{i+1} (t)   \; . \eqno(2.6) $$

Having $i$ as a spin site at time $t+1$ suggests that at time $t$ the
site $i$ was a possible
hopping-particle location site; recall our sublattice rules. We now
identify the random variable
values $\xi_i (t) = 0$ or 1 as the decisions to move the particle to the
right or to the left,
respectively, provided of course that there is a particle at the site
$x=i \ell$ at time $t$. Let us
also define more natural particle occupancy variables,

$$ n_i(t)=[1-m_i(t)]/2 \; , \eqno(2.7) $$

\NI which take on values 0 (empty site) or 1 (occupied site), and
average to $\rho(t)$.

When we calculate $m_i (t+1)$ in terms of the $m_j$ variables at the
preceding time, $t$,
by substituting (2.6) in the $t+1$ version of (2.3), the relation is no
longer linear. The source of
the nonlinearity is the term proportional to $\sigma_{i-2}
\sigma_{i+2} \equiv m_{i-1}m_{i+1}$.
After some algebra, we get, in terms of the $n_i$ variables,

$$ n_i (t+1) = \xi_{i+1} (t) n_{i+1} (t) + \left[ 1 - \xi_{i-1} (t) \right]
n_{i-1} (t)
-2 \xi_{i+1} (t) n_{i+1} (t) \left[ 1 - \xi_{i-1} (t) \right]
  n_{i-1} (t) \; . \eqno(2.8) $$

\NI It is obvious that this nonlinear dynamics corresponds to the
particle-hopping rules as
specified originally. Indeed, the first term adds 1 to $n_i (t+1)$ if
there was a particle at site
$i+1$ at $t$, and if it hopped to the left. The second term adds 1 if
there was a particle at $i-1$
and it hopped to the right. Finally, the last, ``reaction'' term, which is
$(-2)$ times the product of
the first two, ``diffusion'' terms, sets the resulting value to zero if
both diffusion terms were
1, i.e., if two particles hopped to $i$ in the time step $t \to t+1$.
Unfortunately, we found no
other nontrivial quadratic dynamical rules that can be linearized by
the dual-spin
transformation or its simple generalizations. Thus, the reaction
A$+$A$\to$inert is the only
one solvable by the dual-spin method. For instance, if we replace
$(-2)$ by $(-1)$ in (2.8) to have
the reaction scheme  A$+$A$\to$A, then the dual-spin method fails.
The latter reaction can be
solved by another method, however; see [12].

Finally, we note that the hopping rate of a single particle is $H=1$.
Thus, the dimensionless
time variable $\tau$ is given by

$$ \tau = { \left( 1-a^2 \right) t } \; . \eqno(2.9) $$

\

\NI \UN{\bf 3. \ Exact Result for the Particle Density}

The initial particle distribution was selected to correspond to the
random $\pm 1$ assignment
of the initial ($t=0$) spin values. Averaging (2.6) over the stochastic
variables $\xi$ then leads
to the conclusion that $\langle \sigma_i (t) \rangle = 0$ for all times
$t \geq 0$.  The
two-point correlation function, however, is nontrivial. Forming the
product suggested by the
definition (2.5), we obtain, after averaging and some algebra,

$$ G_n (t+1) - G_n (t) = {1-a^2 \over 4} \left[ G_{n-2} (t) - 2 G_n (t) +
G_{n+2} (t) \right] \; .
\eqno(3.1) $$

\NI This relation applies only for $n=2,4,6, \ldots$, and it must be
solved subject to the
boundary condition

$$ G_0 (t \geq 0) = 1 \; , \eqno(3.2) $$

\NI as well as the initial condition

$$ G_{n>0} (0) =0 \; . \eqno(3.3) $$

Let us introduce the generating functions

$$ B_n (v) = \sum_{t=0}^\infty v^t G_n (t) \; , \eqno(3.4) $$

\NI as well as the generating function for the density,

$$ R (v) = \sum_{t=0}^\infty v^t \rho (t) = {1 \over 2} \left[ B_0 (v) -
B_2 (v) \right] \; ,
\eqno(3.5) $$

\NI where the last equality follows from (2.4). Note that we have

$$ B_0 (v) = {1 \over 1-v} \; , \eqno(3.6) $$

\NI while the $n>0$ generating functions satisfy the recursion

$$ {B_n  \over v} - B_n  = {1-a^2 \over 4} \left( B_{n-2}  - 2 B_n  +
B_{n+2}
\right) \; . \eqno(3.7) $$

\NI Each term in (3.7) follows directly from summation of the
appropriate term in (3.1).

This second-order difference equation is quite standard. Only one out
of two independent
solutions behaves regularly for large $n$, in the vicinity of $v=0$.
The solution takes the form

$$ B_n (v) = B_0 (v) \left[ b(v) \right]^{n/2} \; , \eqno(3.8) $$

\NI where the appropriate function $b(v)$ is found directly from
(3.7),

$$ b(v) = 1 + {2 (1-v) \over \left( 1-a^2 \right) v } \left[ 1 - \sqrt{
1+ { \left( 1-a^2 \right) v \over 1-v } } \right] \; . \eqno(3.9) $$

The generating function for the density then follows as

$$ R(v) = {1 \over  \left( 1-a^2 \right) v } \left( { \sqrt{1-a^2 v} \over
\sqrt{1-v} } -1 \right) \;
, \eqno(3.10) $$

\NI where we made several algebraic steps to prepare the resulting
expression for the Taylor
series expansion to yield the density:

$$  \left( 1-a^2 \right) \rho (t) = { (2t+2) ! \over 2^{2t+2} \left[ (t+1) !
\right]^2 }
-\sum_{n=0}^t { (2n)! (2t-2n)! a^{2n+2} \over 2^{2t+1} n! (n+1)!
\left[(t-n)! \right]^2 }
   \;  .  \eqno(3.11) $$

\

\NI \UN{\bf 4. \ Special Limits and Discussion}

In this section we consider three limiting cases in which the solution
(3.11) assumes
less cumbersome forms. First, we note that for the symmetric-
hopping case, $a=0$, only
the first term in (3.11) survives. Since $\tau = t$ for $a=0$, we have

$$ \rho_{a=0} (\tau) = { (2 \tau +2) ! \over 2^{2 \tau +2} \left[ ( \tau
+1) !
\right]^2 } \; . \eqno(4.1) $$

\NI This form provides a convenient analytic continuation to all
$\tau \geq 0$, shown
as the lower solid curve in Figure~1. However, the reader should note
that the actual
time-dependence for $|a| \neq 1$ is always in discrete time steps.

Next, consider the limit $|a| \to 1$, i.e., extreme anisotropy and slow
diffusion.
The ``natural'' time steps $\Delta \tau = (1-a^2) $ then become small
and the
time-dependence approaches a continuous function of $\tau$. As a
result, the
discrete-time difference on the left-hand side of (3.1) can be
replaced by the
derivative $(1-a^2) \partial G_n (\tau) / \partial \tau $. The resulting
equations are
then identical, up to minor notational changes, to those obtained in
the
continuum-time Glauber-dynamics formulation [5,9] of the (isotropic)
reaction
A$+$A$\to$inert. They can be solved, for instance, by the Laplace
Transform method. We only present the final expression for the
density,

$$ \rho_{|a|=1} ( \tau ) = {1 \over 2} {\sl e}^{-\tau/2} \left[ I_0
(\tau/2)  + I_1
(\tau/2) \right] \; , \eqno(4.2) $$

\NI where $I_n$ are the standard Bessel functions. The upper solid
curve in Figure~1
corresponds to (4.2).

Finally, we note that for large times both the temporal and \UN{\sl
spatial\/}
variation of the two-point correlation function can be regarded as
slow and
differences replaced by derivatives. Specifically, the second-difference
expression on
the right-hand side of (3.1) is replaced by the second spatial-coordinate
derivative.
The detailed derivation of this continuum limit and solution for two-point
correlations are presented, e.g., in [11]. In this limit, the hopping
anisotropy is
fully absorbed in the rescaling of time from $t$ to $\tau$. We only
quote the result
for the density

$$ \rho (\tau \gg 1) \simeq {1 \over \sqrt{ \pi \tau } } \; . \eqno(4.3)
$$

\NI This function is shown by the dashed line in Figure~1. It is
asymptotic to the
solid curves for large $\tau$, as well as to the full result (3.11).

The full expression (3.11) is more complicated than the limiting cases
just
discussed. The density, defined for discrete $\tau$-values, ---
integral multiples of
$(1-a^2)$, --- lies between the two limiting solid curves shown in
Figures~1 and 2. For fixed $t$, the $\tau$ values decrease as $|a| \to
1$. Thus, the
crossover from the $a=0$ discrete values to the $|a|=1$ continuous
$\tau$-dependence
occurs for fixed $t$ as shown in Figure~2. The points in
Figure~2 correspond to $t=2$ and

$$ a= 0.00, 0.05, 0.10, \ldots , 0.95 \; ; \eqno(4.4) $$

\NI note that the symbols for the points corresponding to $a=0$ and
$1/20$ largely
overlap in the figure.

In summary, we derived exact results for a two-particle
annihilation reaction with anisotropic hopping on a $d=1$ line. The
main effect of
the anisotropy is to slow down the diffusion-dominated processes,
and it can be
largely absorbed in the redefinition of the time scale. However, in the
regimes where
the discreteness of time steps is nonnegligible, i.e., for short times
and
nearly isotropic hopping, the functional form of the density depends
on the hopping
anisotropy parameter $a$ explicitly. The discrete-time-step reaction
proceeds
slightly faster than the continuous-time reaction, though the
difference is
numerically quite small.

The author wishes to thank M.~Barma, M.L.~Glasser and S.~Nicolis for
helpful comments
and suggestions.

\NP

\centerline{\bf REFERENCES}

{\frenchspacing

\item{1.} M. Bramson and D. Griffeath,
Ann. Prob. {\bf 8}, 183 (1980).

\item{2.} D.C. Torney and H.M. McConnell,
J. Phys. Chem. {\bf 87}, 1941 (1983).

\item{3.} K. Kang, P. Meakin, J.H. Oh and S. Redner,
J. Phys. A{\bf 17}, L665 (1984).

\item{4.} T. Liggett, {\sl Interacting Particle Systems\/}
(Springer-Verlag, New York, 1985).

\item{5.} Z. Racz, Phys. Rev. Lett. {\bf 55}, 1707 (1985).

\item{6.} A.A. Lushnikov, Phys. Lett. A{\bf 120}, 135 (1987).

\item{7.} M. Bramson and J.L. Lebowitz, Phys. Rev. Lett. {\bf 61},
2397 (1988).

\item{8.} Review: V. Kuzovkov and E. Kotomin, Rep. Prog. Phys.
{\bf 51}, 1479 (1988).

\item{9.} J.G. Amar and F. Family, Phys. Rev. A{\bf 41}, 3258 (1990).

\item{10.} L. Braunstein, H.O. Martin, M.D. Grynberg
and H.E. Roman, J. Phys. A{\bf 25}, L255 (1992).

\item{11.} V. Privman, J. Stat. Phys. (1992), in print.

\item{12.} D. ben--Avraham, M.A. Burschka and C.R. Doering,
J. Stat. Phys. {\bf 60}, 695 (1990), and references therein.

\item{13.} V. Privman and M.D. Grynberg, J. Phys. A (1992), in print,
and references therein.

\item{14.} Z. Cheng, J.L. Lebowitz and E.R. Speer, Comm. Pure Appl.
Math. {\bf XLIV}, 971 (1991).

\item{15.} S.A. Janowsky and J.L. Lebowitz, Phys. Rev. A{\bf 45}, 618
(1992).

\item{16.} F.J. Alexander, Z. Cheng, S.A. Janowsky and J.L. Lebowitz, J.
Stat. Phys.
{\bf 68}, 761 (1992).

}

\NP

\centerline{\bf FIGURE CAPTIONS}

\

\NI {\bf Figure~1:\ \ }Particle density as a function of the
dimensionless time
variable $\tau$. Lower solid curve: $a=0$, the actual densities
correspond to integer
$\tau$ values; see Section~4. Upper solid curve: $a=\pm 1$. Dashed
curve: the
$\tau \gg 1$ asymptotic form.

\

\NI {\bf Figure~2:\ \ }The points illustrate the crossover from $a=0$
to $|a|=1$, for
a fixed time, $t=2$. The $a$ values are given in Section~4; see (4.4).
The solid
curves are the same as in Figure~1.

\bye